\documentclass[reprint,nofootinbib,superscriptaddress,amsmath,amssymb,aps]{revtex4-1}
\usepackage{graphicx}
\usepackage{rotating}
\usepackage{dcolumn}
\usepackage{bm}
\usepackage{color}
\usepackage{mathptmx, textcomp}
\usepackage[latin1]{inputenc}
\usepackage{braket}
\usepackage[colorlinks,linkcolor=magenta,anchorcolor=cyan,citecolor=blue]{hyperref}
\usepackage[multiple]{footmisc}
\newcommand{\fig}[1]{Fig.\ref{#1}}
\def\be{\begin{equation}}
\def\ee{\end{equation}}
\def\ba{\begin{eqnarray}}
\def\ea{\end{eqnarray}}

\def\nn{\nonumber}
\def\lf{\left}
\def\rt{\right}

\newcommand{\eq}[1]{(\ref{#1})}

\def\lf{\left}\def\rt{\right}\def\q{\theta}   \def\y {\psi}   \def\p {\pi} \def\a {\alpha} \def\s {\sigma} \def\d {\delta} \def\f {\phi} \def\g {\gamma} \def\h {\eta} \def\j {\varphi} \def\k {\kappa} \def\l {\lambda}  \def\x {\xi} \def\c {\chi}    \def\pd {\partial}\def\p {\pi} \def \inf {\infty}  
\def\Q{\Theta} \def\W{\Omega}       \def\F {\Phi}  \def\L {\Lambda}    \def\grad{\nabla}\def\.{\cdot}
\def\math {\mathcal}
\begin{document}

\title{Holographic Complexity in a Charged Supersymmetric Black Holes}
\author{Jie Jiang}
\email{jiejiang@mail.bnu.edu.cn}
\affiliation{Department of Physics, Beijing Normal University, Beijing 100875, China}
\author{Ming Zhang}
\email{Corresponding author. mingzhang@jxnu.edu.cn}
\affiliation{Department of Physics, Jiangxi Normal University, Nanchang 330022, China}
\date{\today}

\begin{abstract}
For an ordinary charged system, it has been shown that by using the ``complexity equals action'' (CA) conjecture, the late-time growth rate of the holographic complexity is given by a difference between the value of $\Phi_H Q+\Omega_H J$ on the inner and outer horizons. In this paper, we study the influence of the chiral anomaly on the complexity of the boundary quantum system. To be specific, we evaluate the CA holographic complexity of the charged supersymmetric black holes whose bulk action is modified by an additional Chern-Simons term of the electromagnetic fields. As a result, the late-time growth rate of the complexity will be corrected by some additional terms on the inner and outer horizons than the ordinary charged black holes. Our work implies that the late-time growth rate of the complexity can carry the information of the chiral anomaly for the boundary system.

\end{abstract}
\maketitle

\section{Introduction}
In recent years, there has been increasing interest in the idea that apply the quantum information theory into the gravitational theory in the context of the AdS/CFT correspondence. One of the most famous topics in this direction is holographic entanglement entropy. However, it has been claimed that the holographic entanglement entropy is not enough to describe the degrees of freedom inside the horizon of the black hole, where the geometry becomes nonstationary. As a result, the quantum circuit complexity of the boundary quantum system, which is used to describe the minimum number of basic operation gates connected to the target state and the reference state, has been proposed to describes the information inside the black hole horizon \cite{L.Susskind,2}. From the holographic viewpoint, Brown $et\,al.$ suggested that the circuit complexity of the boundary quantum system is given by some classical quantities in the bulk gravitational system, which is called ``holographic complexity''. Then, there are two main conjectures which are proposed to construct this duality. The first one is the ``complexity equals volume" (CV) conjecture \cite{L.Susskind,D.Stanford} and the second one is the ``complexity equals action'' (CA) conjecture \cite{BrL,BrD}. These conjectures have attracted a large number of researchers to study both holographic complexity and circuit complexity in quantum field theory \cite{Jiang:2019pgc,Bhattacharyya:2018bbv,Ali:2018fcz,Ali:2018aon,HosseiniMansoori:2018gdu,Liu:2019smx,Jiang:2019qea,A4,Pan,A2,Guo:2017rul,WYY,Jiang1,
A51,A11,A12,A13,A14,A15,A16,A17,A18,A19,A20,A23,A24,A25,A26,A27,A28,A29,A30,A31,A32,A33,A34,A35,A36,
Fan:2019mbp,Goto:2018iay,JiangL,Nally:2019rnw,Jiang3,Chapman1,Chapman2,Jiang4,SZ,Roberts:2014isa,A37,A38,A39,Jiang1,A2,A3,A4,A5,A6,A7,A8,A9,A10}.

In this paper, we only focus on the second conjecture, i.e., the CA conjecture, which suggests that the circuit complexity of the state $|\y(t_L,t_R)\rangle$ in the boundary quantum system is dual to the on-shell bulk gravitational action $I_\text{WDW}$ within the Wheeler-DeWitt (WDW) patch, which is the causal region that connects the timeslices $t_L$ and $t_R$ on the boundary, i.e., we have
\ba\label{CA}
C_A\lf(|\y(t_L,t_R)\rangle\rt)\equiv\frac{I_\text{WDW}}{\p\hbar}\,.
\ea
It has been argued in Ref. \cite{BrL} that the late-time complexity growth should obey a bound
\ba
\frac{dC}{dt}\leq \frac{2 M}{\p \hbar}\,,
\ea
which is considered to be the Lloyd's bound of boundary quantum system from the viewpoint of the AdS/CFT correspondence \cite{SLloyd}.

For the rotating and charged black holes with multiple horizons,  series works \cite{A2, A4,Pan,Guo:2017rul,WYY,Jiang1,Fan:2019mbp} have showed that the late-time CA complexity growth rate can be expressed as
\ba\begin{aligned}\label{OC}
\lim_{t\to\inf}\frac{dC_A}{d t}&=\frac{1}{\p\hbar}\left[\left(M-\F_H^{(+)}Q-\W_H^{(+)}J\right)\right.\\
&\left.-\left(M-\F_H^{(-)}Q-\W_H^{(-)}J\right)\right]\,,
\end{aligned}\ea
where $Q$ and $J$ are electric charge and angular momentum of the black holes, $\W^{(\pm)}_H$ and $\F_H^{(\pm)}$ are the angular velocity and electric potential of the inner and outer horizons individually. The index $\{(\pm)\}$ represents the quantities related to the outer and inner horizons.  However, all of the above results are obtained in the ordinary charged boundary system. How will things become when we consider an anomalous quantum system? Whether can the complexity carry the information of the anomaly for the boundary system? In this paper, we would like to consider the influence of the chiral anomaly on the complexity. In order to make the boundary system chiral anomaly, it is convenient to introduce an additional Chern-Simons term of the electromagnetic field into the bulk action from the perspective of the AdS/CFT correspondence \cite{Anomaly1}. Then, based on the inflow mechanism, the gauge invariance in the bulk gravitational theory will make the boundary quantum system chiral anomaly \cite{Anomaly1}. Therefore, to study the complexity of the boundary chiral system from the viewpoint of holography, we need a black hole solution in the Einstein-Maxwell-Chern-Simons theory. For simplicity, in this paper, we only consider the minimal gauged five-dimensional supergravity, which is a special case of the Einstein-Maxwell-Chern-Simons theory. The charged rotating black hole solutions in this theory were obtained in \cite{solution1,solution2,solution3,solution4}.

The remaining of this paper is organized as follows. In Sec. \ref{sec2}, we first review the geometry in the charged supersymmetric black holes and introduce some basic spacetime quantities. In Sec. \ref{sec3}, we evaluate the time-dependence and the late-time result of the CA complexity growth rate using the CA conjecture. Finally, the conclusion and discussion are presented in Sec. \ref{sec4}.

\section{Geometry of the five-dimensional charged supersymmetric black holes}\label{sec2}
In this paper, we focus on the five-dimensional minimal gauge supergravity with the bulk action
\ba\begin{aligned}
I_\text{bulk}&=\int_{M}\bm{\epsilon}\left(R+\frac{12}{L^2}-\frac{1}{4}\math{F}\right)\\
&+\frac{1}{3\sqrt{3}}\int_M \bm{F}\wedge\bm{F}\wedge\bm{A} \,,
\end{aligned}\ea
where $\bm{F}=d\bm{A}$ is the electromagnetic strength tensor, $\bm{\epsilon}$ is the volume element of the metric $g_{ab}$, $L$ is the cosmological radius, $R$ is the Ricci Scalar, and we denote $\math{F}=F_{ab}F^{ab}$. Varying the action, the equations of motion read
\ba\begin{aligned}
&R_{ab}-\frac{1}{2}R g_{ab}-\frac{6}{L^2}g_{ab}=\frac{1}{2}\left(F_a{}^cF_{bc}-\frac{g_{ab}}{4}\math{F}\right)\,,\\
&\grad_a F^{ab}=\frac{1}{4\sqrt{3}}\bm{\epsilon}^{bcdef}F_{cd}F_{ef}\,,
\end{aligned}\ea
Generally, the black hole solutions in this gravitational theory carry two angular momenta. For simplicity, in this following, we only focus on the special case which has two equal angular momenta. The solution is given by \cite{solution2}
\ba\begin{aligned}\label{ds2}
ds^2&=-\frac{f}{h}dt^2+\frac{dr^2}{f}+\frac{r^2}{4}(\s_1^2+\s_2^2)+\frac{r^2h}{4}(\s_3-W dt)^2\,,\\
\bm{A}&=-\frac{\sqrt{3}q}{r^2}\left(dt-\frac{j}{2}\s_3\right)\,,
\end{aligned}\ea
in which $\s_1,\s_2,\s_3$ are the usual left-invariant one-fore of the three-sphere and they are given by
\ba\begin{aligned}
\s_1&=-\sin\y d\q+\cos\y\sin\q d\f\,,\\
\s_2&=\cos\y d\q+\sin\y\sin\q d\f\,,\\
\s_3&=d\y+\cos\q d\f\,,
\end{aligned}\ea
and
\ba\begin{aligned}
f&=1+\frac{r^2}{L^2}-\frac{2m(1-\h)}{r^2}+\frac{q^2}{r^4}\left(1-\frac{j^2}{L^2}+\frac{2L^2m\c}{q^2}\right)\,,\\
W&=\frac{2j[(2m-q)r^2-q^2]}{r^6 h}\,,\\
h&=1-\frac{j^2 q^2}{r^6}+\frac{2j^2(m-q)}{r^4}
\end{aligned}\ea
with the quantity
\ba\begin{aligned}
\h=\frac{j^2(m-q)}{m L^2}\,.
\end{aligned}\ea
The electric charge and angular momenta of the black hole are defined by
\ba\begin{aligned}
Q&=\int_{S_\inf}\left(\star \bm{F}-\frac{1}{\sqrt{3}}\bm{F}\wedge \bm{A}\right)\,,\quad J=\int_{S_\inf}\star d\j\,,\\
\end{aligned}\ea
where
\ba
\j^a=\left(\frac{\pd}{\pd\y}\right)^a
\ea
is the axial Killing vector field and $S_\inf$ is a three-sphere at asymptotic infinity. For the charged supersymmetric black hole solution in Eq. \eq{ds2}, the energy, electric charge and angular momentum are given by
\ba\begin{aligned}
M&=\frac{3\W_3 m}{4}\left(1+\frac{\h}{3}\right)\,,\\
Q&=\frac{\sqrt{3}\W_3 q}{4}\,,\\
J&=\frac{\W_3 j(2m-q)}{4}\,,
\end{aligned}\ea
in which $\W_3=16\p^2$ is the volume of the unit three-sphere with the line element
\ba\begin{aligned}
d\W_3^2=d\q^2+\sin^2\q d\f^2+\cos^2\q d\y^2\,.
\end{aligned}\ea

As mentioned in the introduction, in the present work, we only consider the spacetime solution which describes a black hole with two Killing horizons. The inner and outer horizon is determined by $f(r_\pm)=0$. From the line element of this solution, we can see that if there are some regions outside the horizon such that $h(r)\leq 0$, there will exist some naked closed timelike curves lying outside the horizon. Then, the causality of this spacetime will be destroyed. Moreover, if $h(r)<0$ between the inner and outer horizon, the sign of the metric will become $(-,-,-,+,+)$, which is unphysical and cannot be used to describe a spacetime geometry. Taking these into account, in the following, we only focus on the black hole geometry where $h(r)>0$ outside the inner horizon. In this case, the Killing vector fields which generates the inner and outer horizon are given by
\ba\begin{aligned}
k_{(\pm)}^a=\left(\frac{\pd}{\pd t}\right)^a+\W^{(\pm)}_H \left(\frac{\pd}{\pd \y}\right)^a
\end{aligned}\ea
in which $\W^{(\pm)}_H=W(r_{\pm})$ presents the angular velocities of the inner and outer horizons. The corresponding temperature, entropy, and electric potential are given by
\ba\begin{aligned}
&S^{(\pm)}=\frac{\W_3 \p r^3_\pm\sqrt{h(r_\pm)}}{2}\,,\\
T^{(\pm)}=&\frac{f'(r_\pm)}{4\p\sqrt{h(r_\pm)}}\,,\quad \F^{(\pm)}_H=\frac{\sqrt{3}q(2-a\W^{(\pm)}_H)}{2r_\pm^2}\,.
\end{aligned}\ea
\begin{figure*}
\centering
\includegraphics[width=0.95\textwidth]{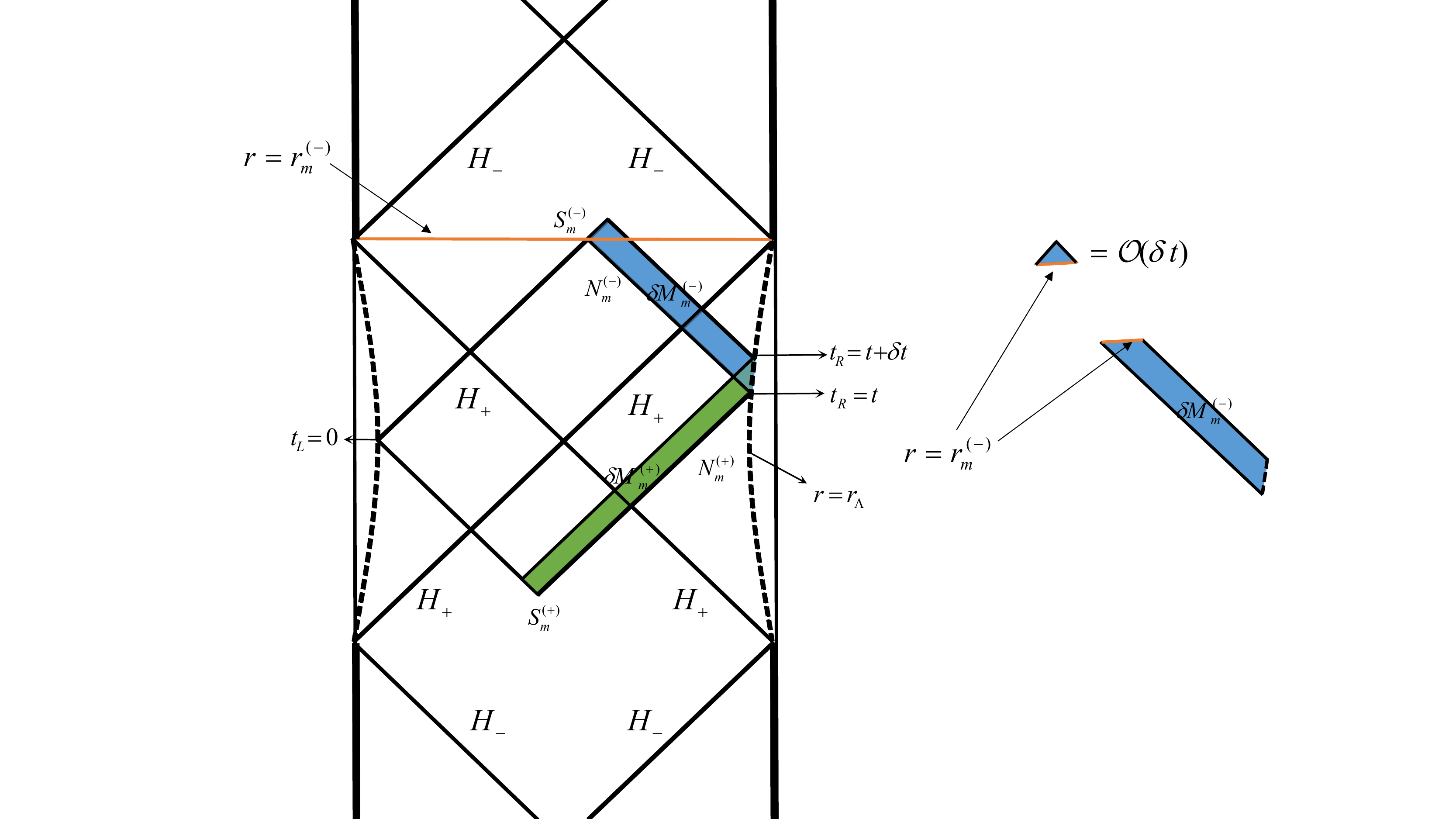}
\caption{ A spacetime diagram of a charged supersymmetric black holes with two Killing horizons. In the right panel, we show the change of the Wheeler-DeWitt patch in this spacetime in which we fix the left boundary time $t_L=0$ and vary the right boundary time $t_R=t$. The dashed lines denote the cut-off surface $r=r_\L$ at asymptotic infinity to regulate the divergence near the AdS boundary. In the right panel, we show that the bulk region $\d M^{(-)}_m$  of the WDW patch can be generated by the Killing vector $k_{(-)}^a$ through $N_m^{(-)}$ under the first-order approximation of $\d t$.}\label{WDW}
\end{figure*}

\section{complexity growth rate in CA conjecture}\label{sec3}

In this section, we would like to evaluate the growth rate of the holographic complexity for the above charged supersymmetric black holes based on the CA conjecture. That is to say, we need to evaluate the on-shell action within the WDW patch. As suggested by Lehner $et\, al.$ in \cite{A3}, the total on-shell action should include not only the bulk action but the surface terms, joint terms, and counterterms as well. Here the surface terms and joint terms are added to ensure the variational principle well-posed when the bulk region has a non-smooth boundary. The counterterms are introduced to make the total action invariant under the reparametrization of the null generator on the null segment. According to Ref. \cite{A3}, the total action can be expressed as
\ba\begin{aligned}\label{fullaction}
&I=I_\text{bulk}+2\int_Bd^4x\sqrt{|h|}K
\pm2\int_Sd^3\q\sqrt{\s}\h\\
&+2\int_Nd\l d^3\q \sqrt{\g} \k +2\int_Nd\l d^3\q\sqrt{\g}\Q\ln\lf(\ell_{\text{ct}}\Q\rt)\,,
\end{aligned}\ea
where $B$ and $N$ are the non-null and null segments of the boundary $\pd M$ separately, $S$ is a three-dimensional joint of the non-smooth boundary. Here $h_{ab}$ and $K$ are the induced metric and the trace of the extrinsic curvature on the non-null surface $B$, $\h$ is a transformation parameter on the joint $S$, $\g_{ab}$ is the induced metric on the cross section of the null segment $N$, $\l$ is the parameter of the null generator $k^a$ for the null segment $N$, the parameter $\k$ is given by
$k^a\grad_ak^b=\k k^b$ and it measures the failure of the parameter $\l$ to be an affine parameter for the null generator $k^a$, $\Q=\nabla_ak^a$ is the expansion scalar of the null generator, and $\ell_\text{ct}$ is some arbitrary constant parameter.

In \fig{WDW}, we show the change of the WDW patch in the charged supersymmetric black holes which has two Killing horizons. Considering that the spacetime is invariant under the shift transformation $t_R\to t_R-\d t, t_L\to t_L+\d t$, without loss of generality, we fix the left boundary time at $t_L=0$ and only vary the right boundary time $t_R=t$ of the WDW patch. Since we only focus on the growth rate of the total action within the WDW patch, we can ignore the higher-order corrections $\math{O}(\d t)$ of $\d t$. From \fig{WDW}, we can see that the bulk region $\d M^{(\pm)}_m$ can be generated by the horizon Killing vector $k_{(\pm)}^a$ through the null segment $N_m^{(\pm)}$ under the first-order approximation of $\d t$ (Generally, it can be generated by the Killing vector $k_{(\pm)}^a+C \j^a$ with any constant $C$ because the null segment $N_m^{(\pm)}$ is invariant under the axial transformation). Moreover, we shall choose $\l$ to be an affine parameter of the null generators. As a result, the surface term vanishes on all null segments. With these in mind, the action change we need to evaluate can be divided into
\ba\begin{aligned}
\d I_\text{WDW}=I_{{\d M}^{(-)}_m}-I_{\d M^{(+)}_m}+\d I_{S_m^{(-)}}-\d I_{S_m^{(+)}}+\d I_\text{ct}\,.
\end{aligned}\ea

\noindent\textbf{Bulk contributions}

We start by calculating the bulk contributions from the region $\d M_m^{(\pm)}$. Because the calculations are same for $\d M_m^{(-)}$ and $\d M_m^{(+)}$, we will neglect the index $(\pm)$ for all the quantities. From the equations of motion, we can further obtain
\ba\begin{aligned}
R&=\frac{1}{12}\math{F}-\frac{20}{L^2}\,,\\
R_{ab}&=\frac{1}{2}F_a{}^cF_{bc}-g_{ab}\left(\frac{\math{F}}{12}+\frac{4}{L^2}\right)\,.
\end{aligned}\ea
Then, the bulk contribution from the region $\d M_m^{(\pm)}$ gives
\ba\begin{aligned}\label{bulkgrav}
&I_{\d M_m}=-\left(\frac{8}{L^2}+\frac{\math{F}}{6}\right)\int_{\d M}\sqrt{-g}d^5x+\frac{1}{3\sqrt{3}}\int_{\d M}\bm{F}\wedge\bm{F}\wedge\bm{A}\\
&=-\d t\left(\frac{8}{L^2}+\frac{\math{F}}{6}\right)\int_{N}\star \x+\frac{\d t}{3\sqrt{3}}\int_{N}\x\.\left(\bm{F}\wedge\bm{F}\wedge\bm{A}\right)\,.\\
\end{aligned}\ea
Considering the facts that $k^a$ is a Killing vector and ${R^a}_{[bcd]}$ vanishes, we have
\ba\begin{aligned}\label{grad}
\grad_a \grad^a k^b&=-{R^b}_a k^a\\
&=\left(\frac{\math{F}}{12}+\frac{4}{L^2}\right)k^b-\frac{1}{2}k^aF_{ac}F^{bc}\,.
\end{aligned}\ea
For the second term, we have
\ba\begin{aligned}
k^aF_{ac}F^{bc}&=k^a(d \bm{A})_{ac}F^{bc}\\
&=\math{L}_{k}\bm{A}_c F^{bc}+\grad_c \F F^{bc}\\
&=-\grad_a\left(\F F^{ab}\right)+\F\grad_a F^{ab}\\
&=-\grad_a\left(\F F^{ab}\right)+\frac{\F}{4\sqrt{3}}\bm{\epsilon}^{bcdef} F_{cd}F_{ed}\,,
\end{aligned}\ea
where we have denoted
\ba\begin{aligned}
\F^{(\pm)}=-A_ak^a_\pm\,.
\end{aligned}\ea
Then, we can get
\ba\begin{aligned}
2\grad_a \grad^a k^b=\left(\frac{\math{F}}{6}+\frac{8}{L^2}\right)k^b+\grad_a\left(\F F^{ab}\right)-\frac{\F}{4\sqrt{3}}\bm{\epsilon}^{bcdef} F_{cd}F_{ed}
\end{aligned}\nn\\\ea
Using the language of differential forms, the above identity can be expressed as
\ba\begin{aligned}
-\left(\frac{\math{F}}{6}+\frac{8}{L^2}\right)\star\bm{k}=d(\star d\bm{k}-\F \star \bm{F})+\frac{\F}{\sqrt{3}}\bm{F}\wedge\bm{F}
\end{aligned}\ea
On the other hand, we have
\ba\begin{aligned}
k\.\left(\bm{F}\wedge\bm{F}\wedge\bm{A}\right)
&=2(k\.\bm{F})\wedge\bm{F}\wedge\bm{A}-\F\bm{F}\wedge\bm{F}\\
&=2d\F\wedge\bm{F}\wedge\bm{A}-\F\bm{F}\wedge\bm{F}\\
&=2d(\F\bm{F}\wedge\bm{A})-3\F\bm{F}\wedge\bm{F}
\end{aligned}\ea
Summing the above results, we can further obtain
\ba\begin{aligned}
\frac{I_{\d M_m}}{\d t}&=\int_{N}d\left(\star d \bm{k}-\F \star \bm{F}+\frac{2}{3\sqrt{3}}\F \bm{F}\wedge\bm{A}\right)\\
&=\int_{\pd N}\left(\star d \bm{k}-\F \star \bm{F}+\frac{2}{3\sqrt{3}}\F \bm{F}\wedge\bm{A}\right)\\
&=\int_{S_\inf}\star d \bm{k}-\int_{S_m}\star d \bm{k}+\F(r_m)\int_{S_m} \left(\star \bm{F}-\frac{2}{3\sqrt{3}}\bm{F}\wedge\bm{A}\right)\\
&=\int_{S_\inf}\star d \bm{\x}+\W_H J+\F(r_m)Q-\c(r_m)\F(r_m)+\math{P}(r_m)\,.
\end{aligned}\ea
where we have denoted
\ba\begin{aligned}
\c(r)&=-\frac{1}{3\sqrt{3}}\int_{S_r}\bm{F}\wedge \bm{A}=\frac{\W_3q^2 a^2}{4\sqrt{3}r^4}\,,\\
\math{P}^{(\pm)}(r)&=-\int_{S_r}\star d \bm{k}_\pm\\
&=\frac{\W_3 r^3}{32h}[4h f'-4f h'+h^3r^2W'(\W_H^{(\pm)}-W)]\,,
\end{aligned}\ea
in which $S_r$ is a three-sphere with radius $r$. After completing all of the indexes, the bulk contributions yield
\ba\begin{aligned}
&\frac{d I_\text{bulk}}{d t}=\frac{I_{\d M_m^{(-)}}}{\d t}-\frac{I_{\d M_m^{(-)}}}{\d t}\\
&=\left(\W^{(-)}_H -\W^{(+)}_H\right)J+\left[\F^{(-)}(r_m^{(-)})-\F^{(+)}(r_m^{(+)})\right]Q\\
&-\left[\c(r_m^{(-)})\F^{(-)}(r_m^{(-)})-\c(r_m^{(+)})\F^{(+)}(r_m^{(+)})\right]\\
&+\math{P}^{(-)}(r_m^{(-)})-\math{P}^{(+)}(r_m^{(+)})
\end{aligned}\ea
At the late times $t=t_R\to \inf$, we have $r_m^{(\pm)}\to r_\pm$. Then, the growth rate of the bulk action becomes
\ba\begin{aligned}
\lim_{t\to\inf}&\frac{d I_\text{bulk}}{d t}=\left(\W^{(-)}_H J +\F^{(-)}_H Q-\c_H^{(-)}\F_H^{(-)}+T^{(-)}S^{(-)}\right)\\
&-\left(\W^{(+)}_H J +\F^{(+)}_H Q-\c_H^{(+)}\F_H^{(+)}+T^{(+)}S^{(+)}\right)
\end{aligned}\ea
where we have denoted
\ba\begin{aligned}
\c_H^{(\pm)}= \c(r_\pm)\,.
\end{aligned}\ea
\begin{figure*}
\centering
\includegraphics[width=0.49\textwidth]{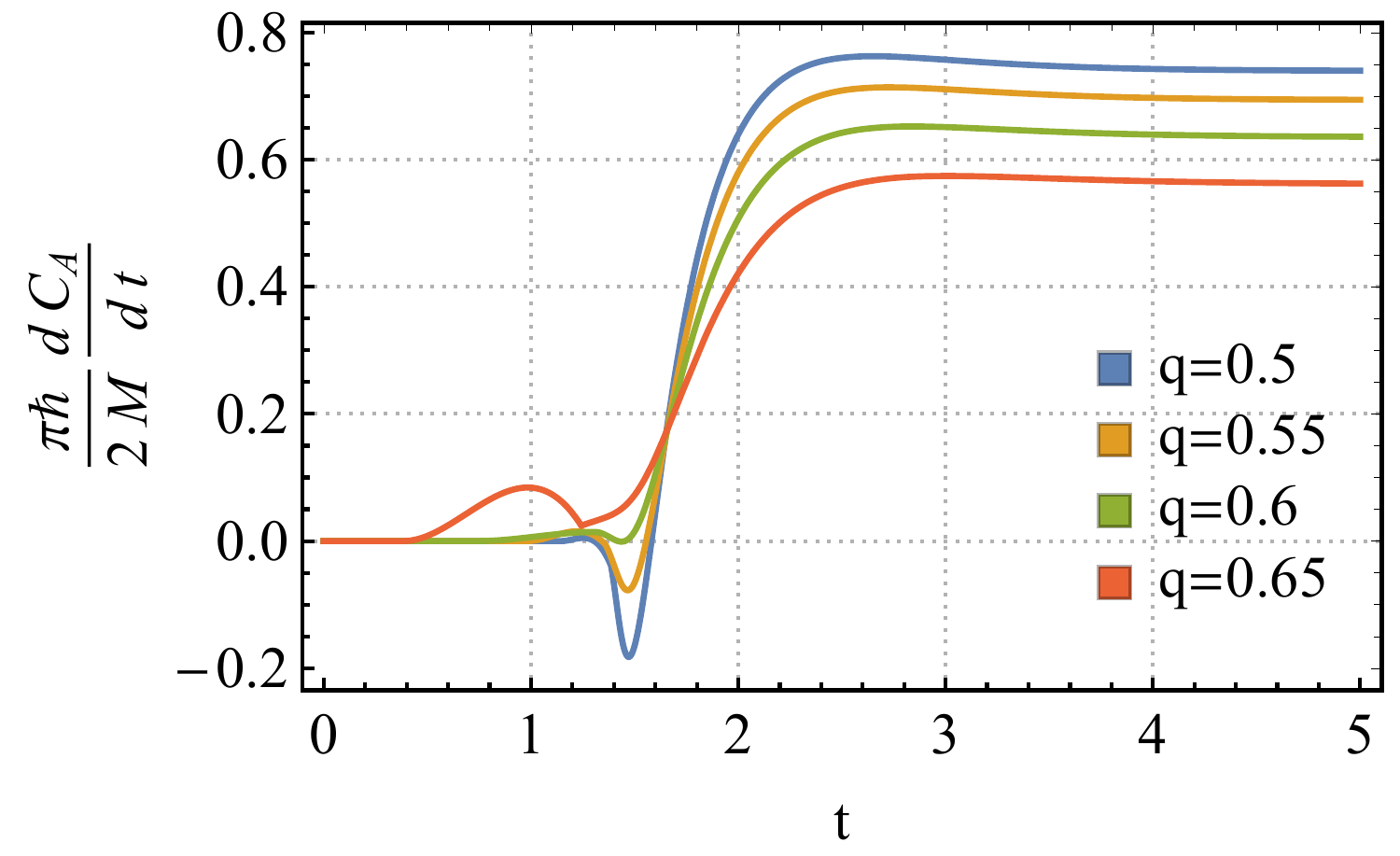}
\includegraphics[width=0.49\textwidth]{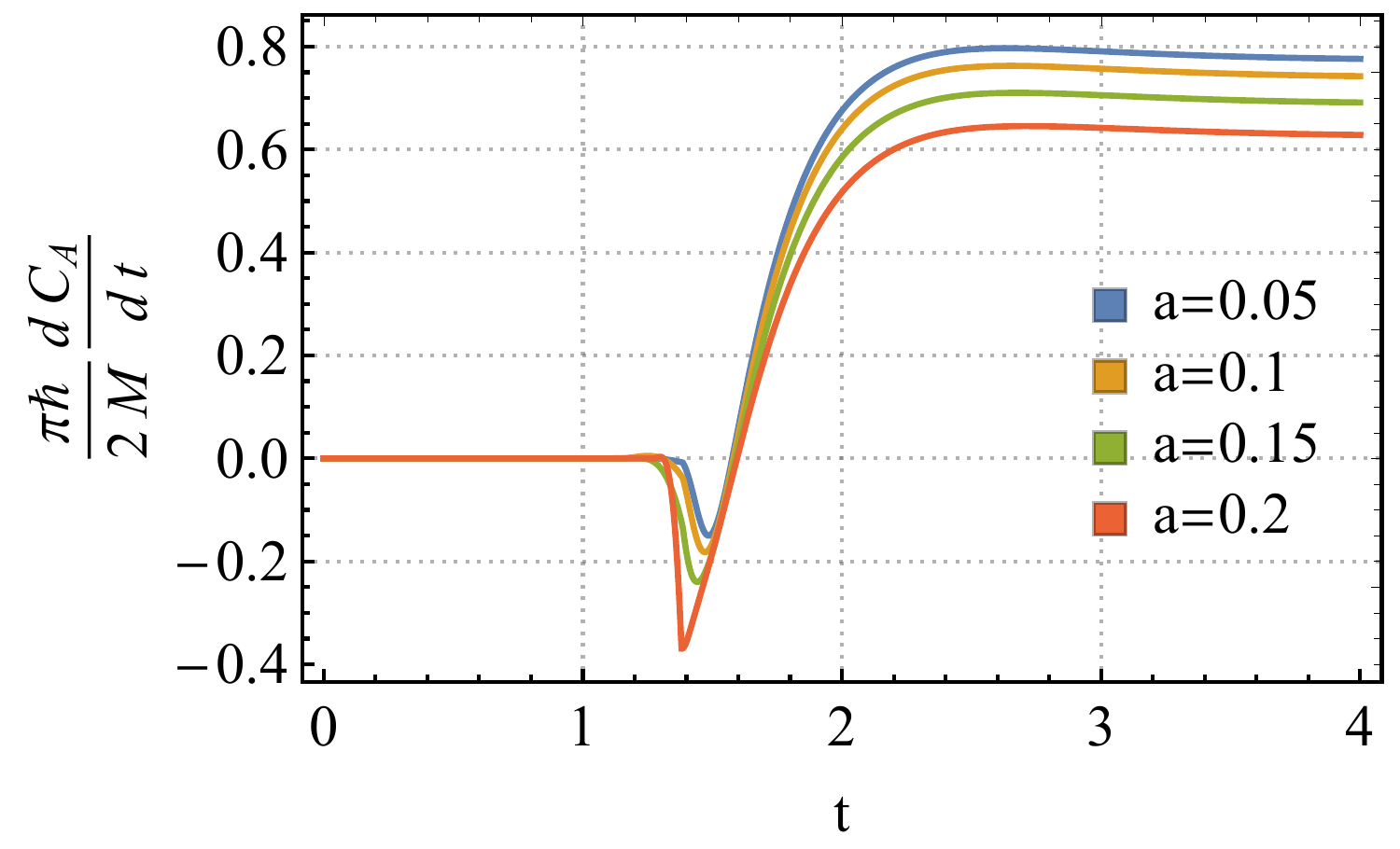}
\caption{Plots show the time-dependence of the complexity growth rate. In the left panel, we vary the charge parameter $q$ and fix $m=1, L=1, l_\text{ct}=0.5, a=0.1$. In the right panel, we vary the angular momenta parameter $a$ and fix $m=1, L=1, l_\text{ct}=0.5, q=0.5$.}\label{fig}
\end{figure*}

\noindent \textbf{Joint contributions}

We next turn to calculate the joint contributions from meeting points $S_m^{(\pm)}$. We first focus on the joint $S_m^{(+)}$ which is formed by the intersection of the past right and past left null segments. From the line element \eq{ds2}, it is not hard to see
\ba\begin{aligned}\label{k1k2}
(k_1)_a&=(dv)_a+\frac{1}{\tilde{f}(r)}(dr)_a\,,\\
(k_2)_a&=(dv)_a-\frac{1}{\tilde{f}(r)}(dr)_a\,
\end{aligned}\ea
with
\ba\begin{aligned}
\tilde{f}(r)=\frac{f(r)}{\sqrt{h(r)}}
\end{aligned}\ea
are the affinely null generator of the past right and past left null segments separately. The transformation parameter of this joint is given by $\h=\ln(|k_1\cdot k_2|/2)$ \cite{A3}. Using the above null generators, we can further obtain
\ba\begin{aligned}
k_1\cdot k_2=-\frac{2h(r)}{f(r)}\,.
\end{aligned}\ea
Then, the joint contribution from $S_m^{(+)}$ is given by
\ba\begin{aligned}
I_{S_m^{(+)}}=-\left.\frac{\W r^3\sqrt{h}}{4}\ln \lf[-f/h\rt]\right|_{r=r_m^{(+)}}\,.
\end{aligned}\ea
From Eq. \eq{k1k2}, we can see that these two segments are determined by $t+r^\star(r)=t_R$ and $t-r^\star(r)=t_L=0$. Here $r^\star(r)$ is the tortoise coordinate and it is defined by
\ba\begin{aligned}\label{rstar}
r^\star(r)&=-\int_r^\inf \frac{dr}{\tilde{f}(r)}\,,
\end{aligned}\ea
where we have fixed the coordinate to satisfy the condition
\ba\begin{aligned}
\lim_{r\to \inf}r_\star (r)=0\,.
\end{aligned}\ea
From Ref. \cite{A4}, the tortoise coordinate can also be written as
\ba\begin{aligned}
r^\star(r)&=\frac{\ln(|r-r_+|/r)}{g(r_+)(r_+-r_-)}-\frac{\ln(|r-r_-|/r)}{g(r_-)(r_+-r_-)}\\
&-\frac{1}{r_+-r_-}\int_{r}^{\inf} G(r)dr
\end{aligned}\ea
with
\ba\begin{aligned}
g(r)&=\frac{\tilde{f}(r)}{(r-r_+)(r-r_-)}\,,\\
G(r)&=\frac{g(r_+)r-g(r)r_+}{g(r_+)g(r)r(r-r_+)}
-\frac{g(r_-)r-g(r)r_-}{g(r_-)g(r)r(r-r_-)}\,.
\end{aligned}\ea
From the above discussion, the radius of $S_m^{(+)}$ can be further obtained,
\ba\label{rmp}
r^\star(r_m^{(+)})=-\frac{t}{2}\,.
\ea
Then, we can get the its change rate as
\ba\label{drp}
\frac{d r_m^{(+)}}{d t}=-\frac{1}{2}\tilde{f}(r_m^{(+)})\,.
\ea
Using the above results, we have
\ba\begin{aligned}
\frac{d I_{S_m^{(+)}}}{dt}=\left.\frac{\W_3 r^3(h f'-f h')}{8h}+\frac{\W_3r^2(6h+r h')f\ln [-f/h]}{16h}\right|_{r_m^{(+)}}\,.
\end{aligned}\nn\\\ea
Similarly, the contribution from the joint $S_m^{(-)}$ can also be obtained with a same calculation. Then, the growth rate of the joint contribution is given by
\ba\begin{aligned}
\frac{d I_\text{joint}}{dt}=\left.\frac{\W_3 r^3(h f'-f h')}{8h}+\frac{\W_3r^2(6h+r h')f\ln [-f/h]}{16h}\right|^{r_m^{(+)}}_{r_m^{(-)}}\,,
\end{aligned}\nn\\\ea
where we have used the result
\ba\label{rmm}
r^\star(r_m^{(-)})=\frac{t}{2}\,.
\ea
for the meeting point $S_m^{(-)}$. At the sufficient late times, we have
\ba\begin{aligned}
\lim_{t\to \inf}\frac{d I_\text{joint}}{dt}=T^{(+)}S^{(+)}-T^{(-)}S^{(-)}\,.
\end{aligned}\ea
We can see that this term will cancel the $TS$ term in the bulk contributions.
\\

\noindent\textbf{Counterterm contributions}

Finally, we are going to evaluate the counterterm contributions. We first consider the past right null segment. From Eq. \eq{k1k2}, the expansion of the null generator on this segment gives
\ba\begin{aligned}
\Q=\frac{6h+r h'}{2r \sqrt{h}}\,.
\end{aligned}\ea
Then, the counterterm of the past right null segment can be shown as
\ba\begin{aligned}
I_\text{ct}^\text{(pr)}=\left.\frac{\W_3}{8}\int_{r_m^{(+)}(\l)}^{r_\L(\l)}d\l r^2(6h+r h')\ln\lf(\frac{(6h+rh')\ell_\text{ct}}{2r\sqrt{h}}\rt)\right|_{r=r(\l)}\,,
\end{aligned}\nn\\\ea
Here $\l$ is the affine parameter of the null generator $k_1^a=(\pd/\pd\l)^a$. Together with Eq. \eq{k1k2}, we can find that $r'(\l)=\sqrt{h}$, which implies that $d\l=dr/\sqrt{h}$. Then, we have
\ba\begin{aligned}
I_\text{ct}^\text{(pr)}=\frac{\W_3}{8}\int_{r_m^{(+)}}^{r_\L}dr \frac{r^2(6h+r h')}{\sqrt{h}}\ln\lf(\frac{(6h+rh')\ell_\text{ct}}{2r\sqrt{h}}\rt)\,.
\end{aligned}\ea
Using Eq. \eq{rmp}, we can further obtain
\ba
\frac{d I_\text{ct}^\text{(pr)}}{dt}=\left.\frac{\W_3r^2(6h+r h')f}{16h}\ln\lf(\frac{(6h+rh')\ell_\text{ct}}{2r\sqrt{h}}\rt)\right|_{r_m^{(+)}}
\ea
Again, the counterterm contributions from other segments can also be obtained similarly and the final result can be expressed as
\ba
\frac{d I_\text{ct}}{dt}=\left.\frac{\W_3r^2(6h+r h')f}{8h}\ln\lf(\frac{(6h+rh')\ell_\text{ct}}{2r\sqrt{h}}\rt)\right|^{r_m^{(+)}}_{r_m^{(-)}}\,.
\ea
\newpage
\noindent\textbf{Complexity growth rate}

Summing all the previous results and using the CA conjecture in Eq. \eq{CA}, one can further obtain
\ba\begin{aligned}\label{dCdt}
&\p \hbar\frac{d C_A}{d t}=\left(\W^{(-)}_H -\W^{(+)}_H\right)J+\left[\F^{(-)}(r_m^{(-)})-\F^{(+)}(r_m^{(+)})\right]Q\\
&-\c(r_m^{(-)})\F^{(-)}(r_m^{(-)})+\c(r_m^{(+)})\F^{(+)}(r_m^{(+)})\\
&+\tilde{\math{P}}^{(-)}(r_m^{(-)})-\tilde{\math{P}}^{(+)}(r_m^{(+)})\\
&-\left[\frac{\W_3r^2(6h+r h')f}{16h}\ln\lf(-\frac{(6h+rh')^2f\ell_\text{ct}^2}{4r^2h^2}\rt)\right]^{r_m^{(-)}}_{r_m^{(+)}}\,,
\end{aligned}\nn\\\ea
where we have denoted
\ba\begin{aligned}
\tilde{\math{P}}^{(\pm)}(r)=\frac{\W_3 h^2r^5W'}{32}(\W_H^{(\pm)}-W)\,.
\end{aligned}\ea

In \fig{fig}, we show the time-dependence of the CA complexity growth rate in the charged supersymmetric black holes. This figure shows a similar behavior with the case of the RN-AdS black holes where the late time value is approached above.

Finally, we consider the late-time limit of the complexity growth rate. From Eq. \eq{dCdt}, it is easy to get
\ba\begin{aligned}
\lim_{t\to\inf}\frac{d C_A}{d t}&=\frac{1}{\p\hbar}\left[\left(\W^{(-)}_H J +\F^{(-)}_H Q-\c_H^{(-)}\F_H^{(-)}\right)\right.\\
&\left.-\left(\W^{(+)}_H J +\F^{(+)}_H Q-\c_H^{(+)}\F_H^{(+)}\right)\right]\,.
\end{aligned}\ea
The above results show the difference from the ordinary charged system as shown in Eq. \eq{OC}. Here the late-time growth rate will be corrected by some additional terms which are evaluated on the inner and outer horizons, i.e., $\c_H^{(+)}\F_H^{(+)}-\c_H^{(-)}\F_H^{(-)}$.

\section{Conclusion and Discussion}\label{sec4}

In this paper, we considered the five-dimensional minimal gauged supergravity, which is a special case of the Einstein-Maxwell-Chern-Simons theory. From the perspective of AdS/CFT, the dual bound system of this bulk gravity is a real anomaly. To study the influence of the arial anomaly of the boundary system to the complexity, we evaluated the growth rate of the holographic complexity in a charged and rotating supersymmetric black holes by using the CA conjecture. As a result, we found that the time-dependence of the complexity growth rate shares similar behavior as the cases of the RN black holes. However, the late-time rate is different from the result \eq{OC} of the ordinary charged system. Here it is corrected by an additional term $\c_H\F_H$ which is evaluated on the inner and outer horizons. These imply that the late-time growth rate of the complexity carries some information of the chiral anomaly for the boundary quantum system. Moreover, from the above calculations, it is not hard to verify that the additional corrections $\a (\c_H^{(+)}\F_H^{(+)}-\c_H^{(-)}\F_H^{(-)})$ will also appear in a general Einstein-Maxwell-Chern-Simons gravity, which includes a general Chern-Simons term
\ba\begin{aligned}
I_\text{CS}=\frac{\a}{2\sqrt{3}}\int_{M}\bm{F}\wedge\bm{F}\wedge\bm{A}\,.
\end{aligned}\ea
with a coupling constant $\a$. Our work strongly implies that the anomaly of the boundary will play an important role in complexity.

\section*{acknowledgements}
Jie Jiang  is supported by the National Natural Science Foundation of China (Grants No. 11775022 and
11873044). Ming Zhang is supported by the Initial Research Foundation of Jiangxi Normal University with Grant No. 12020023.

\end{document}